\documentclass[sn-nature]{sn-jnl}

\usepackage{graphicx}%
\usepackage{multirow}%
\usepackage{amsmath,amssymb,amsfonts}%
\usepackage{amsthm}%
\usepackage{mathrsfs}%
\usepackage[title]{appendix}%
\usepackage{xcolor}%
\usepackage{textcomp}%
\usepackage{manyfoot}%
\usepackage{booktabs}%
\usepackage{algorithm}%
\usepackage{algorithmicx}%
\usepackage{algpseudocode}%
\usepackage{listings}%

\newcommand{\araa}{Annu. Rev. Astron. Astrophys.}   
\newcommand{\apj}{Astrophys. J.}   
\newcommand{\apjl}{Astrophys. J. Lett.}   
\newcommand{\apjs}{Astrophys. J. Suppl. Ser.}   
\newcommand{\ao}{Appl. Opt.}   
\newcommand{\aap}{Astron. Astrophys.}   
\newcommand{\aapr}{Astron. Astrophys. Rev.}   
\newcommand{\jgr}{J. Geophys. Res.}   
\newcommand{\mnras}{Mon. Not. R. Astron. Soc.}   
\newcommand{\nat}{Nature} 
\newcommand{\prd}{Phys. Rev. D}   
\newcommand{\prl}{Phys. Rev. Lett.}   

\begin{document}

\title[PeVatrons]{The hunt of PeVatrons as the origin of the most energetic photons observed in our Galaxy}

\author*[1]{\fnm{Emma} \sur{de O\~na Wilhelmi}}\email{emma.de.ona.wilhelmi@desy.de}
\author*[2]{\fnm{Rub\'en} \sur{L\'opez-Coto}}\email{rlopezcoto@iaa.es}


\author[3,4]{\fnm{Felix} \sur{Aharonian}}

\author[5,6]{\fnm{Elena} \sur{Amato}}

\author[7,8,9]{\fnm{Zhen} \sur{Cao}}

\author[10]{\fnm{Stefano} \sur{Gabici}}

\author[3]{\fnm{Jim} \sur{Hinton}}
\affil[1]{\orgname{Deutsches Elektronen-Synchrotron DESY}, \orgaddress{\street{Platanenallee 6}, \city{Zeuthen}, \postcode{D-15738}, \country{Germany}}}

\affil[2]{\orgname{Instituto de Astrof\'isica de Andaluc\'ia, CSIC}, \orgaddress{\city{Granada}, \postcode{18008}, \country{Spain}}}

\affil[3]{\orgname{Max-Planck-Institut f\"ur Kernphysik}, \orgaddress{\street{P.O. Box 103980}, \city{Heidelberg}, \postcode{D-69029}, \country{Germany}}}

\affil[4]{\orgname{Dublin Institute for Advanced Studies}, \orgaddress{\street{31 Fitzwilliam Place}, \city{Dublin}, \country{Ireland}}}

\affil[5]{\orgname{INAF Osservatorio Astrofisico di Arcetri}, \orgaddress{\street{Largo E. Fermi 5}, \city{Firenze}, \postcode{I-50125}, \country{Italy}}}

\affil[6]{\orgname{Università degli Studi di Firenze}, \orgaddress{\street{Via Sansone 1}, \city{Sesto Fiorentino}, \postcode{I-50019}, \country{Italy}}}

\affil[7]{\orgname{Key Laboratory of Particle Astrophysics \& Experimental Physics Division \& Computing Center, Institute of High Energy Physics, Chinese Academy of Sciences}, \orgaddress{\street{19B Yuquan Road, Shijingshan District}, \city{Beijing}, \country{China}}}
\affil[8]{\orgname{TIANFU Cosmic Ray Research Center}, \orgaddress{\street{No. 1500, Kezhi Road, Tianfu New Area}, \city{Chengdu, Sichuan}, \country{China}}}

\affil[9]{\orgname{University of Chinese Academy of Sciences}, \orgaddress{\street{19 Yuquan Road, Shijingshan District}, \city{Beijing}, \country{China}}}
\affil[10]{\orgname{Universit\'{e} de Paris, CNRS, Astroparticule et Cosmologie},\orgaddress{\street{10 Rue Alice Domon et Leonie Duquet},\city{Paris}, \postcode{F-75006}, \country{France}}}


\abstract{Ultrarelativistic particles called cosmic rays permeate the Milky Way, propagating through the Galactic turbulent magnetic fields. The mechanisms under which these particles increase their energy can be reasonably described by current theories of acceleration and propagation of cosmic rays. There are, however, still many open questions as to how to reach petaelectronvolt (PeV) energies, the maximum energy believed to be attained in our Galaxy, and in which astrophysical sources (dubbed {\it PeVatrons}) this ultra-high energy acceleration happens. In this article, we describe the theoretical conditions for plasma acceleration to these energies, and the Galactic sources in which these conditions are possible. These theoretical predictions are then confronted with the latest experimental results, summarising the state-of-the-art of our current knowledge of PeVatrons. We finally describe the prospects to keep advancing the understanding of these elusive objects, still unidentified more than one hundred years after the discovery of cosmic rays.

}

\maketitle

\section{Introduction}
\label{sec:intro}

\vspace{3mm}

\begin{figure}
    \centering
    \includegraphics[width=0.8\textwidth]{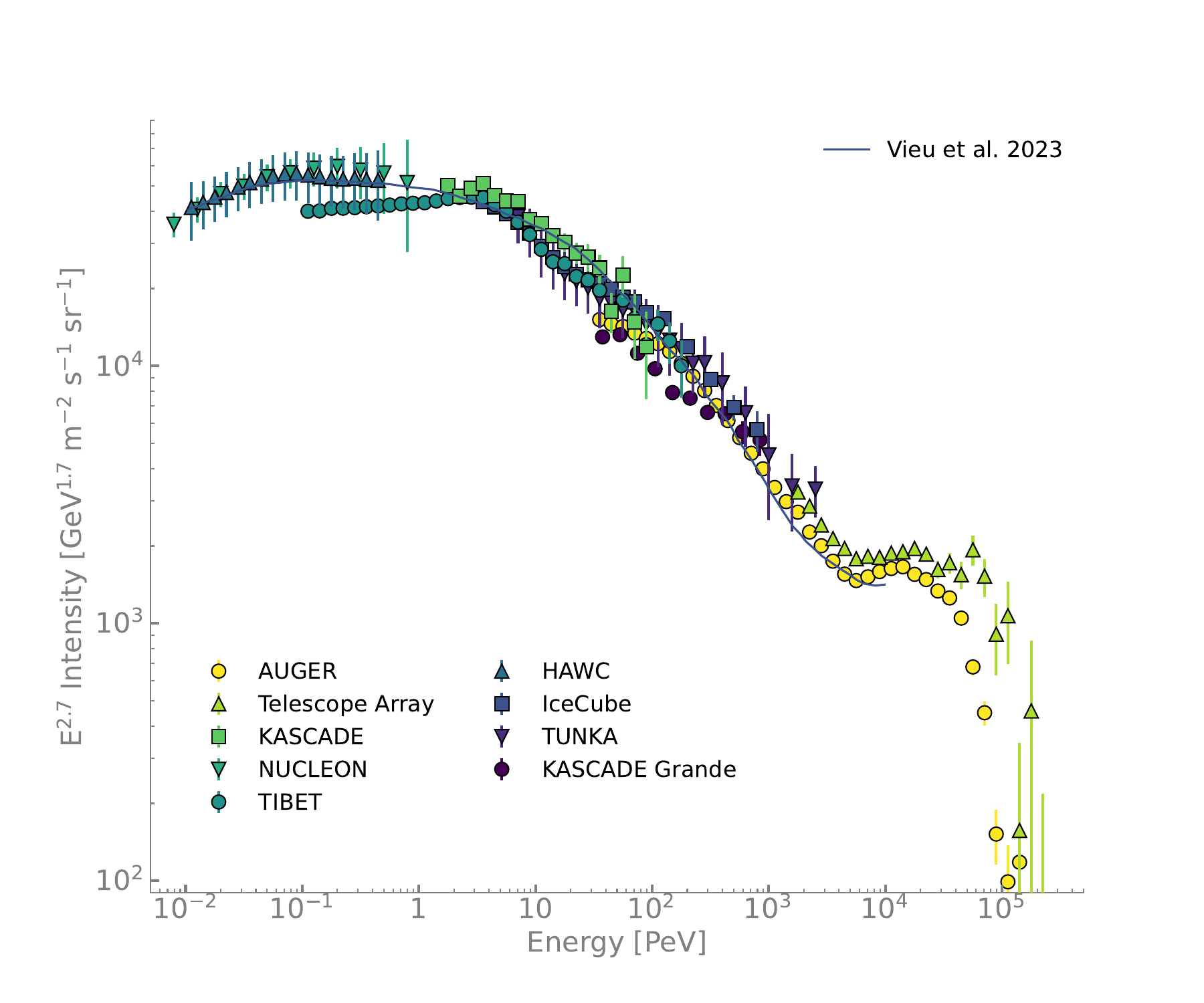}
    \caption{Cosmic Ray all-particle spectrum measured by different experiments in the region above 5 TeV.
    Measurements from KASCADE \citep{DataAP_Kascade}, Tunka \citep{DataAP_Tunka}, TIBET \citep{DataAP_Tibet}, Telescope Array \citep{DataAP_TA}, NUCLEON \citep{DataAP_Nucleon}, KASCADE-Grande \citep{DataAP_KascadeGrande}, IceCube \citep{DataAP_Icecube}, HAWC \citep{DataAP_Hawc} and AUGER \citep{DataAP_Auger} are represented. Error bars represent 1-sigma statistical errors in the data points. The black line shows the theoretical expectation calculated by \citep{Vieu2023}, assuming contribution from accelerated cosmic rays in different regions associated with massive stellar clusters. Figure adapted from \citep{The_CR_spectrum}.}
    \label{fig:spectrum}
\end{figure}

The identification of astronomical sources contributing to the locally observed relativistic particles called Cosmic Rays (CRs) has historically been and still is one of the most important unsolved questions of high-energy astrophysics. The extension of the energy spectrum of CRs well beyond PeV energies (see Fig. \ref{fig:spectrum}) and the break in the spectrum present at these energies indicate the presence of PeVatrons  - {\it non-hypothetical} (undoubtedly existing) extreme particle accelerators in the Milky Way (for recent reviews, see \citep{2013A&ARv..21...70B, 2014IJMPD..2330013A, 2019IJMPD..2830022G}). PeVatrons, i.e. acceleration sites in our galaxy in which CRs (composed mainly by protons) are accelerated up to at least PeV energies, can be linked to more than one source population. In this regard, comprehensive multi-wavelength (from radio to gamma-rays) and multi-messenger (neutrinos and CRs) studies of PeVatrons are essential for understanding their nature. Yet, unambiguous searches for hadronic PeVatrons and their studies can be conducted only by Ultra-High Energy (UHE) ($\geq 100$~TeV) gamma-rays and neutrinos - the secondary (neutral \& stable) products of interactions of primary CRs in the PeV. In both $pp$ and $p\gamma$ collisions, a sizeable part of the energy of the accelerated proton is transferred to secondary gamma rays and neutrinos. Although the exact fraction depends on the spectrum of the primary particle and, in the case of photo-meson interactions, also on the spectrum of the target photons, in both processes, approximately 10 percent of the primary energy is released in the form of secondary gamma rays and neutrinos through the production and decay of intermediate neutral and charged mesons. Thus, detecting UHE gamma rays from an astronomical object would tell us that the energy of the parent proton exceeds 1 PeV, i.e. we deal with an hadronic PeVatron. Regarding UHE gamma-rays produced by leptons, the only realistic 
mechanism at energies above 100 TeV is the Inverse Compton (IC) Scattering on the 2.7~K Cosmic Microwave Background (CMB) photons \citep{1996MNRAS.278..525A,2010A&A...523A...2M}. In the energy interval between 30\,TeV to 3\,PeV, when the Klein-Nishina effect becomes progressively important, the relation between the energy of a gamma-ray and the primary electron can be approximated, with good accuracy (better than 10\%), as  $E_{\rm e} \approx  0.37 (E_\gamma/100 \, \rm TeV)^{0.77}  \, \rm PeV$ \citep{2021Sci...373..425L}. Thus the 100\,TeV {\it threshold}, defined as the low-energy edge of the UHE domain, can be set as the signature of both hadronic and leptonic PeVatrons.

In the Very-High Energy (VHE) ($\geq 100$~GeV) domain, the great success in the 2000s of the Stereoscopic Systems of Imaging Atmospheric Cherenkov Telescopes (IACTs) like e.g. HEGRA, H.E.S.S., MAGIC, VERITAS dramatically changed the status of the field. The detection of more than 250 sources of TeV gamma-ray emission belonging to more than a dozen galactic and extragalactic source populations elevated the field to the level of a modern astronomical discipline. However, the ground-based gamma-ray astronomy results related to PeVatrons have been somehow ambiguous. These days, we are witnessing a new revolution - this time in the UHE domain. The recent fascinating discoveries by Wide Field-of-view (WFoV) gamma-ray detectors such as HAWC, Tibet AS Gamma, and LHAASO collaborations of UHE gamma-ray sources with energy spectra extending beyond 100 TeV opened a new (UHE) window in the cosmic electromagnetic spectrum. To a large extent, the report of more than forty galactic gamma-ray sources by LHAASO above 100 TeV provides solid observational material for studying the nature of PeVatrons \citep{LHAASO_catalog}. The majority of these objects have potential counterparts from different source catalogs, but only a few of them are currently associated with known astronomical objects. As indicated above, all UHE gamma-ray sources are related, in one way or another, to leptonic and hadronic PeVatrons.  

In the following, we describe in Sec. "PeV Accelerators: theoretical considerations" the astrophysical conditions to accelerate particles to PeV energies, in Sec. "Ultra high energy gamma-ray sources" the latest results towards the identification of PeV accelerators in the very-high-energy (VHE) and UHE regimes, and in the last section, the prospects and conclusion statements.

\vspace{5mm}

\section{PeV Accelerators: theoretical considerations}
\label{sec:particle_acceleration}

\begin{figure}
    \centering
    \includegraphics[width=0.8\textwidth]{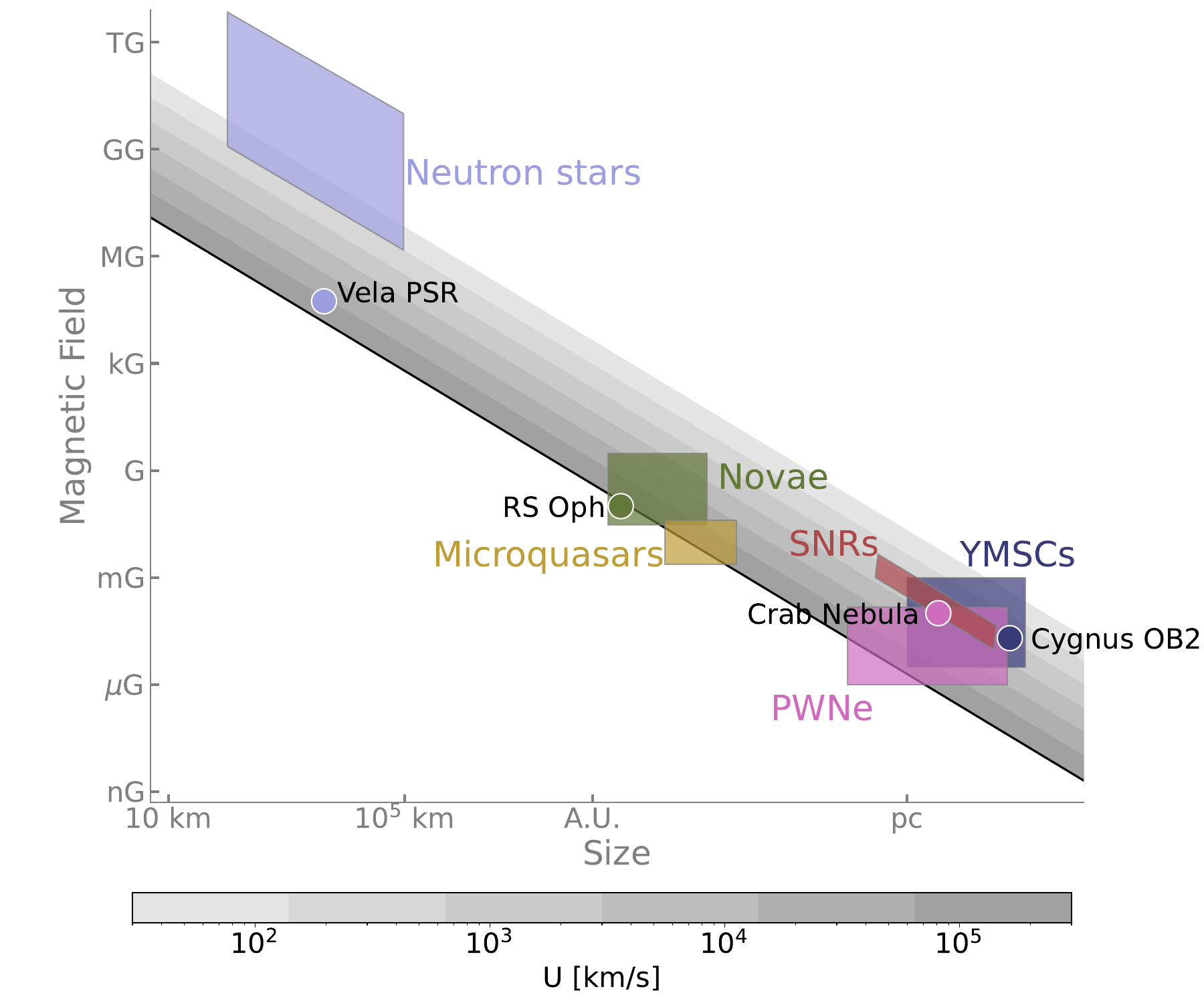}
    \caption{Magnetic field versus size of potential Galactic PeVatrons. The figure, known as the \textit{Hillas Plot} is adapted from \cite{gernot_maier_2022_6037985} (see reference therein for the order of magnitude used for the Galactic objects included). A few remarkable TeV-bright gamma-ray sources are displayed as guidance. The gray area shows the minimum magnetic field and size required to reach 1\,PeV for several outflow velocities U (see Eq. \ref{eq:hillas}), with the lower boundary corresponding to ultra-relativistic outflows (U$\sim$c)}  \label{fig:hillas_plot}
\end{figure}

The theoretical identification of astrophysical regions in which CRs can be accelerated to PeV energies requires the understanding of the acceleration and transport of such CRs, considered as a plasma, permeating the Galaxy \citep{2018AdSpR..62.2731A, 2021JPlPh..87a8401A}.
The most effective way to accelerate a particle of electric charge $Ze$ ($e$ being the elementary charge) is to inject it into a region of space filled with a uniform and static electric field $\vec{\cal E}$.
The particle would be accelerated along $\vec{\cal E}$ by the electric force and, after traveling a distance $L$, would acquire an energy $E_{max} = Z e {\cal E} L$. However, astrophysical plasmas are typically highly conductive, and therefore unable to support large-scale electric fields - electric charges would move under the effect of such fields and quickly restore neutrality \cite{spitzer}. On the other hand, magnetic fields $\vec{\cal B}$ are ubiquitous in the Universe.
With the magnetic field comes the inducted electric field $\vec {\cal E} = (\vec U/c)\times \vec {\cal B}$ with $\vec U$ the characteristic local velocity of plasma motion (in the relativistic case expressed as $\vec U = \vec \beta c$).   
Substituting this into the expression for $E_{max}$ given above, one obtains the following relation between $E_{\rm max}$ and the properties of the system in terms of size $L$, typical velocity $U$ and magnetic field $B$:
\begin{equation}
\label{eq:hillas}
E_{max}^{th} = Z \left( \frac{e}{c} \right) L~ U~ {\cal B} \sim  0.3~ Z \left( \frac{L}{\rm pc} \right) \left( \frac{U}{1000~{\rm km/s}} \right) \left( \frac{\cal B}{100~\mu{\rm G}} \right)  \rm PeV\ .
\end{equation}
This is known as the Hillas criterion \cite{hillas1,hillas2}. The Hillas criterium adapted to Galactic sources is shown in Fig. \ref{fig:hillas_plot}. Several types of source populations, bright in the TeV regime, are included for reference, along with a few of the most archetypal individual sources with their estimated size and magnetic field magnitudes. The colored area shows the maximum energy reached for different values of flow speed $U$.

An equivalent way to express this is to state that particles are accelerated at the fastest possible rate allowed by electrodynamics: $r_{acc}^{max} = (1/E)(dE/dt) = Ze{\cal B} c/E$.
More generally, in the case of {\bf diffusive} acceleration occurring at a {\bf quasi-parallel} shock propagating at speed U=$\beta c$, even in the most optimistic case of Bohm diffusion when particles scatter on average every Larmor radius $r_L$, the acceleration rate r$_{acc}$ = U$^2$/D$_{\rm B}$ (being D$_{\rm B}$=1/3 $r_L$c the Bohm diffusion coefficient) is reduced to \cite{2014IJMPD..2330013A}: 
\begin{equation}
r_{acc}=\frac{3 \beta^2 Ze {\cal B}c}{E}\ .
\label{eq:racc}
\end{equation}

This reduction is critical because in most situations the maximum energy an accelerator can provide is far lower than the theoretical value $E_{max}^{th}$ in Eq.\ref{eq:hillas}, and determined by the competition between the rate of energy gain and losses or by the lifetime of the accelerator, through the condition $r_{acc}(E_{max}) =\max(r_{loss},1/t_{life})$. It should be noted that acceleration can be faster, in principle, at quasi-perpendicular non-relativistic shocks \citep{1994JGR....9919351G}, where shock drift (SDA), rather than diffusive shock acceleration (DSA) can take place. Recent numerical studies have shown that SDA can be effective for ions \citep{2023PhRvL.131i5201O} if the shock Mach number is sufficiently high. Whether leptons can also be accelerated to energies beyond a few GeV in such shocks is not yet clear.

For hadrons losses are typically negligible (their large mass makes them ineffective radiators) and it is then clear that the maximum achievable energy increases with the typical velocity of the accelerator and with the strength of the magnetic field it hosts. These ingredients can be used in a straightforward way to identify the sources that can be responsible for the acceleration of PeV CRs. Supernova Remnants (SNRs) are the earliest suggested candidates (since 1930s'), and the association finds observational support in the presence, in young SNRs, of electrons with an energy of tens of TeV and largely amplified magnetic fields, in the $100 \mu$G range \cite{vink2012}, that could result from the development of instabilities induced by accelerated CRs leaving the source \cite{bell2013}. A magnetic field in this range, combined with a shock velocity $\sim 10^4$km/s would lead to an acceleration to PeV energies within a few tens of years (see green area in Fig. \ref{fig:hillas_plot}). No signs of PeV particles have been found however in gamma-ray observations of young SNRs (i.e. no gamma-ray emission has been detected at tens of TeV, corresponding to the radiation of PeV particles) \cite{2021JPlPh..87a8401A,2021Univ....7..324C}, but theoretical modeling suggests that the PeVatron phase lasts for a very short time ($\lesssim 100$yrs), and likely occurs only for a very small fraction of remnants resulting from especially energetic explosions \cite{Cristofari20}. In this case, rather than detecting UHE gamma rays from a SNR, it could be easier to detect emission from escaped PeV particles interacting with some dense cloud at some distance from the accelerator \cite{gabici2007}. Alternative scenarios have been recently suggested, in which a soft hadron spectrum can be formed \cite{2011MNRAS.418.1208B,2019MNRAS.488.2466B,2019ApJ...881....2M,2019ApJ...885...11H,2020ApJ...905....2C,2022ApJ...925...48X}. This soft scenario might also explain the lack of gamma-ray detection from young SNRs \cite{2023arXiv231017082C}. 

In recent years, the detection of gamma rays of a number of Young Massive Star Clusters (YMSC) (e.g. \cite{aharonian2019}) revived the interest in these sources as possible CR hadronic PeVatrons.
Particle acceleration can take place either at the star cluster wind termination shock \cite{morlino2021} or in the interior of the wind- and supernova-blown superbubble that forms around the cluster \cite{vieu2022}.
While in the former case, particles undergo diffusive shock acceleration, in the latter particle energization is due to an interplay of acceleration at SNR shocks and stochastic re-acceleration in the strongly turbulent plasma (see \cite{bykov2014} and references therein). The magnetic fields and shock velocities in these shocks are, in general, lower than for SNRs, but large energies can be achieved given the longer live time ($t_{life}$ $\sim$ Myr) available for acceleration.    
An additional scenario for particle acceleration to PeV energies and beyond is offered by supernovae exploding in the magnetized wind of massive star clusters \cite{vieu2022b}. This idea is supported by results from recent numerical simulations suggesting that the ambient magnetic field might be significantly amplified in the inner region of a star cluster due to wind-wind collisions \cite{badmaev2022}. 

Also, pulsars and Pulsar Wind Nebulae (PWNe) are potentially able to accelerate particles to PeV energies, with relativistic plasma flows ($U\approx c$). Protons, or even heavier ions, can in principle be extracted from a pulsar, just as well as primary electrons \cite{1980ApJ...235..576C}. In terms of number, however, they will be much fewer protons than electrons and positrons, after pair production has taken place in the magnetosphere. In the pulsar magnetosphere, actual charge starvation electric fields develop along the magnetic field where particles can be accelerated subject only to curvature losses (which affect both hadrons and leptons). For as much as the pulsar physics might be complicated, the ingredients that play a role in acceleration (magnetic field and extent of the acceleration region) combine in such a way that the maximum accelerating potential for particles that leave the pulsar magnetosphere only depends on the star rotational energy loss rate $\dot E$ \cite{1969ApJ...157..869G}, and turns out to be
\cite{2022ApJ...930L...2D}:
\begin{equation}
    E_{\rm max}^{\rm PSR}=Z e \sqrt{\dot E/c}\ .
    \label{eq:emaxpsr}
\end{equation}

Particles leave the pulsar magnetosphere as part of a highly relativistic magnetized wind, which terminates at a shock when the final spectrum of relativistic particles that makes PWNe bright non-thermal sources is believed to be finally formed. Interestingly, it has been shown \cite{2022ApJ...930L...2D} that also at the termination shock (the other powerful accelerator in the pulsar-PWN system) the maximum achievable energy is the same as in the magnetosphere, and described by Eq.\ref{eq:emaxpsr}. When losses cannot be ignored, as is very often the case for electrons, the maximum energy of particles is determined by the equilibrium between the acceleration and the energy loss rates: $r_{acc} = r_{loss}$. Losses are usually dominated by synchrotron radiation in the ambient magnetic field unless the latter is parallel to the accelerating electric field or contains an energy density that is below that of the CMB radiation. The first situation requires the plasma to be charge-starved to ensure the persistence of the electric field: the main astrophysical instance is found in pulsar magnetospheres, discussed above. The latter scenario requires a magnetic field below $3 \mu G$, which is uncommon in high-energy particle accelerators, and in any case implies the possibility of reaching very high energy only for very long-lasting systems, according to Eq.\ref{eq:racc}. If the dominant loss mechanism is synchrotron radiation, then the rate of energy losses is $r_{loss}^{sync}=(\sigma_T c/6 \pi)B^2 E/(m_e c^2)^2$, with $m_e$ the electron mass and $\sigma_T$ the Thomson cross-section. The condition $r_{acc}\le r_{loss}$ defines:
\begin{equation}
E_{\rm max}^{\rm loss}\approx 30 \left(\frac{U}{10^3 \rm km/s}\right) \left( \frac{\cal B}{100~\mu{\rm G}} \right)^{-1/2} \rm TeV\ , 
\label{eq:emaxloss} 
\end{equation}
where $U$ and $\cal B$ have been scaled to values typical of young SNRs, while the energy would be larger by a factor of a few for electrons accelerated in a YMSC.
 Eq.\ref{eq:emaxloss} shows that acceleration of electrons to PeV energies requires either relativistic flow speed or a speed of several $10^4$~km/s and $B<10 \mu$G. While the latter condition might apply to some YMSCs \cite{Vieu2023}, the first condition certainly applies to pulsars and PWNe and makes these sources primary leptonic PeVatrons. Thus, lepton acceleration to PeV energies and beyond seems to be possible only at the termination shock of PWNe around pulsars with sufficiently large $\dot E$, namely $\dot E>2 \times 10^{35} \rm erg/s$ \cite{2022ApJ...930L...2D} and magnetic fields at the termination shock that do not exceed 3\,mG (see Eq. 5 in \citep{2022ApJ...930L...2D}), or in star clusters with fields below $10 \mu \rm G$. Other source populations are excluded when combining the acceleration rate described above, with the condition $r_{acc}<1/T_{life}$ (with $T_{life}$ the lifetime of the system) and Eq.\ref{eq:emaxloss}. SNRs are automatically excluded as lepton accelerators and so is the Galactic Center region.


\section{Ultra high energy gamma-ray sources}
\label{sec:ExperimentalResults}

In the last years, a large fraction of the sky has been investigated using deep observations to uncover the origin of the PeV CRs. Among these observations, surveying large regions of the Galactic plane is probably the best strategy to study the population of PeV accelerators in an unbiased manner. The first surveys of the Galactic plane above $\sim$0.5\,TeV were done with arrays of IACTs \citep{2002A&A...395..803A,2018A&A...612A...1H,2015ICRC...34..750P}, being the most sensitive and extensive among them the one done by H.E.S.S. (the H.E.S.S. Galactic Plane Survey, HGPS), covering from l = 250 to 65 deg and latitudes
$\vert b \vert \le 3$ deg. Despite the large variety and richness of the results derived from the HGPS, the sensitivity of IACTs is limited beyond $\gtrsim$ 50\,TeV, and thus, it is non-optimal to investigate PeVatrons. The reason is intrinsic to the detection technique in the VHE domain: the high-energy photons are absorbed in the atmosphere and re-emitted in the form of a shower of secondary particles. These charged particles radiate Cherenkov light which can be collected by the large reflectors of Cherenkov Telescopes, providing an excellent description of the photon direction, energy, and nature. Showers of particles produced by photons with energies beyond tens of TeV are larger and penetrate deeper, becoming more difficult to detect. This requires telescopes spread out in extensive regions with large field-of-view cameras (see Sec. "Prospects and Final Remarks"). Contrarily, the detection of gamma rays with WFoV instruments becomes extremely efficient in the UHE regime. The use of water pools or tanks covering large areas, and particle detectors sampling the arrival time, nature (via muon detectors), and density distribution of the showers provide an improvement in sensitivity by at least one order of magnitude at 50\,TeV with respect to IACTs, reaching more than 2 orders of magnitude above 100\,TeV (see Fig. \ref{fig:sensitivities}). This impressive performance has resulted in the largest step forward in our understanding of the PeV sky. The first catalog published by the LHAASO collaboration reports the results obtained during the first $\sim$500\,days of observations \cite{LHAASO_catalog}. From the 90 sources listed in the catalog, 43 are detected at energies beyond 100 TeV at more than 4$\sigma$ significance level. The origin of these gamma-ray sources is yet unclear but given the extended nature of most of the sources (ranging from $\sim0.2^{\rm o}$ to $>2^{\rm o}$, \cite{2023arXiv231017082C}) and their location lying on the Galactic plane, they surely represent a new population of PeV accelerators in our Galaxy. 

Among the bright sources shining up to PeV energies, the Crab Nebula stands out as an extreme accelerator, reaching~1.4 PeV with a spectrum showing gradual steepening over three energy decades \cite{2021Sci...373..425L}. However, the Crab pulsar does not seem to be the only pulsar powering a PeVatron in the sky: 35 of the sources listed in the LHAASO catalogs are spatially coincident with high spin-down luminosity pulsars \cite{LHAASO_catalog}, and 10 out of the 12 sources reaching PeV energies \cite{2021Natur.594...33C} can also be potentially associated to pulsars \cite{2022ApJ...930L...2D}. PWNe constitute thus a distinct class of PeV electron accelerators. However, the acceleration of electrons to PeV energies is a theoretical challenge, given the strong synchrotron losses that PeV electrons suffer at these energies as discussed in the previous section. Understanding the operation of these perfectly designed machines accelerating electrons at a rate close to the absolute theoretical limit should be a principal goal of future studies. 

The identification of the origin of the gamma-ray emission is not always easy, particularly in the case of moderately extended gamma-ray regions ($\sim0.5^{\rm o}$), due to the limited angular resolution of WFoV detectors ($\sim0.4^{\rm o}$ above 3 TeV \cite{2021ChPhC..45b5002A}). This can be solved using complementary IACT arrays observations to resolve the source morphology (at lower energies), with an angular resolution of $\le 0.1^{\rm o}$ \citep{2009arXiv0901.2153F}. The two techniques thus complement each other: whereas WFoVs can be used to find the regions in which the PeV accelerators are located, IACTs can pinpoint the direction where the PeV particles are accelerated within the larger region constrained by WFoV and resolve the gamma-ray morphology, which can also be used to describe the particle propagation around the accelerator. In the case of diffuse, large sources, WFoVs have a clear advantage with respect to the current Cherenkov telescopes, given the large field-of-view and duty cycle, which translates into a similar sensitivity in large regions in the sky. For instance, the gamma-ray emission observed surrounding the stellar cluster Cygnus OB2 \citep{Li:2021icb, 2021NatAs...5..465A, 2011Sci...334.1103A} extends to 1.4 PeV from a region up to several degrees, suggesting that the latter hosts a super-PeVatron boosting the energy of hadrons beyond 10 PeV \citep{2019NatAs...3..561A, 2021NatAs...5..465A, Li:2021icb}. This very extended gamma-ray emission can be spatially resolved, revealing an intensity profile for protons that follows a R$^{-1}$ distribution, with R the distance to the cluster. IACTs can resolve regions closer to the clusters, and a similar radial profile was obtained for Westerlund~1 \citep{2012A&A...537A.114A, 2022A&A...666A.124A}, where the photon spectrum seems to extend also to $\sim$100\,TeV without a visible softening.

The idea that YMSCs might be primary contributors to VHE CRs has lately taken some relevance due to the surprising lack of evidence of PeV emission coming from young, fast-shock supernova remnants, the traditional population ascribed to the origin of the PeV CRs \cite{2019IJMPD..2830022G}. CRs are indeed found in several evolved SNRs such as W44 or IC 433 \citep{2013Sci...339..807A}. But in these SNRs, the shock has already slowed down and the gamma-ray spectrum decays before reaching 1\,TeV. In addition, the emission is possibly due to reaccelerated, rather than freshly accelerated, CRs (see e.g. \citep{2016A&A...595A..58C}). It is premature, however, to draw a verdict. Because of the steep $\gamma$-ray spectra, detecting $\geq 100$~TeV $\gamma$ rays is challenging even for the LHAASO detectors. On the other hand, some sources from the first LHAASO Catalogue are likely linked to molecular clouds located in the vicinity of middle-aged SNRs. This can most naturally be explained as $\gamma$-ray "echoes"  of PeVatrons operating at the early stages of the evolution of these SNRs.
A similar scenario was invoked to explain the gamma-ray emission observed from the Galactic center ridge, which was the first hadronic source detected with a particle spectrum extended at least up to 0.4\,PeV (95\% cl) \cite{2016Natur.531..476H, 2018A&A...612A...9H, 2020A&A...642A.190M}. The gamma-ray emission was attributed to CRs escaping from a central accelerator in the Galactic center region. The identification of this central accelerator is not trivial though: the center of our Galaxy hosts many objects capable of producing CRs of high energy, including the supermassive black hole, but also a SNR, a PWN, and some of the most massive YMSCs in our Galaxy. 

Deeper observations with instruments on the south hemisphere with better angular resolution will provide an answer to this question (see Sec. "Prospects and Final Remarks").


\section{Prospects and Final Remarks }
\label{sec:prospects}

A new boost in the UHE regime demands a dramatic improvement in terms of future instrumentation. These developments need to be focused on the key aspects identified in this document: we need to improve the resolution of the measurements and increase the number of sources and energy coverage, that is, improving angular and energy resolution, as well as sensitivity. 

Deep studies of currently known sources can still shed light on the PeV acceleration problem. Improvements in analysis methods from current IACTs may also give extra insights thanks to improved resolution (e.g. \citep{Parsons}) and high energy background rejection power \citep{Shilon, Olivera1, Olivera2}. 
Improvements in datasets of extended air shower arrays like that from Pass 5 from HAWC \citep{HAWC:2023amj, pass5} or the inclusion of the outrigger array that significantly improves the high energy sensitivity of the array \citep{hawc_outriggers}.

\begin{figure}
    \centering
    \includegraphics[width=0.8\textwidth]{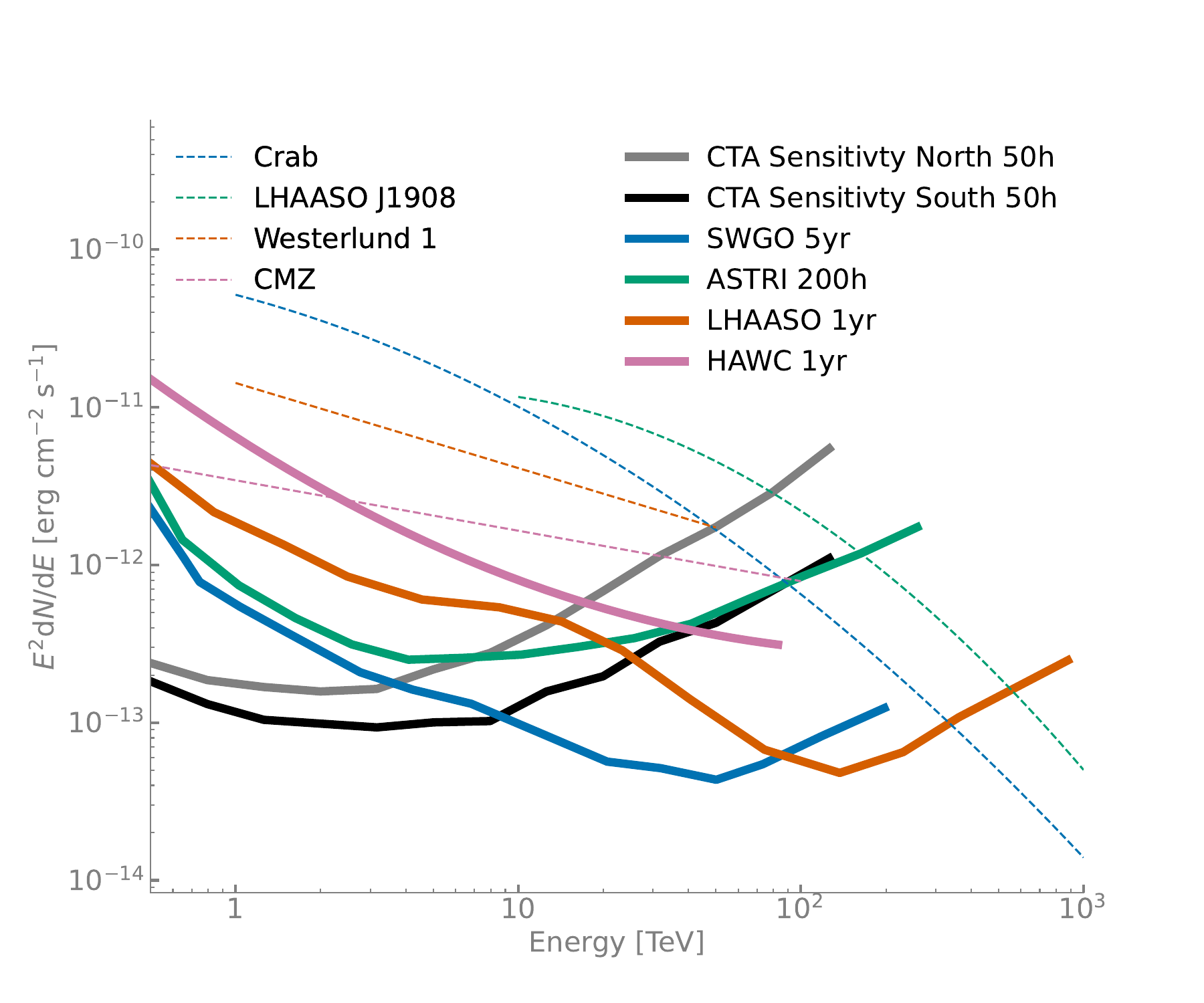}
    \caption{Sensitivity of different instruments with spectra of selected UHE sources. Sensitivities extracted from \citep{fermi_pass8, 2016APh....72...76A, HESS:2015cyv, 2017ApJ...843...39A, 2019arXiv190208429A, 2019arXiv190502773C, cherenkov_telescope_array_observatory_2021_5499840, 2022icrc.confE.884L}. Spectra from the Crab Nebula \citep{2021Sci...373..425L}, Westerlund 1 \citep{2022A&A...666A.124A}, LHAASO J1908 \citep{2021Natur.594...33C} and CMZ \citep{2016Natur.531..476H}. Note that in the case of SWGO the curve provided is the so-called 'strawman' sensitivity. The goal performance is significantly better, especially in the UHE domain~\cite{swgo_icrc}.
    }
    \label{fig:sensitivities}
\end{figure}

Future observations, especially those performed in the Southern Hemisphere where most of the Galactic plane is accessible, with improved sensitivity, angular, and energy resolution, will help in solving the question of PeV acceleration in the Galaxy. We illustrate the sensitivity of different current and future observatories in Fig. \ref{fig:sensitivities}, compared to the spectra of selected UHE sources. Observatories like CTAO will allow us to study the morphology of Supernova Remnants \citep{2023APh...15002850A}, star-forming regions like Westerlund 1, and the spectrum and morphology of the Galactic Centre (GC) region emission. The Southern Wide-field Gamma-ray Observatory (SWGO,\cite{swgo_icrc}) aims for LHAASO-like highest energy sensitivity in the Southern hemisphere, together with unprecedented angular resolution for a wide-field gamma-ray instrument. SWGO can help extend the spectra of accelerators into the UHE domain and probe emission on large angular scales more easily than CTA.
LHAASO and HAWC will continue collecting data in the Northern Hemisphere, with improvements such as the  HAWC outriggers that will improve the measurements currently being performed. Other arrays currently being constructed like the Large Array of Cherenkov Telescopes Array (LACT) as an extension of LHAASO experiment IACT instruments \citep{2021EPJC...81..657A} and the ASTRI Mini-Array in the Northern hemisphere \citep{ASTRI_PeVatrons} or ALPACA \citep{alpaca_status} and its future upgrades \citep{mega_alpaca} in the Southern one, will also play an important role in PeVatron studies. 

In addition to these efforts in the gamma-ray regime, PeVatrons will also become a science prime topic in multiwavelength and multimessenger astronomy.
Observations using observatories like IceCube \citep{icecube} and KM3Net \citep{km3net}, in the search for coincident neutrinos with candidate PeV sources are crucial to unequivocally associate the gamma-ray emission with hadronic mechanisms in the accelerator. Likewise, the synchrotron counterpart of the radiation process in such magnetized regions (either from high energy PeV primary electrons or secondary ones from hadronic interactions) is expected in the X-ray domain \cite{2022ExA....54...23P}.

In conclusion, in the last years, observations using satellites such as {\it Fermi}-LAT and IACTs provided many clues about the acceleration and transport of CRs in our Galaxy. GeV/TeV monitoring of our Galaxy has revealed many particle accelerators, both leptonic and hadronic. Moreover, the detection of hadronic signatures in a handful of molecular-cloud interacting SNRs endorsed the latter as hadronic CRs accelerators at least to a few TeV \cite{2013Sci...339..807A,2011ApJ...742L..30G,2016ApJ...816..100J,2019A&A...623A..86A}. Still, despite the long search, the origin of Galactic CRs with the highest energies, up to PeV or beyond, remains inconclusive. This is however changing fast and we are closer than ever to finally uncover the origin of Galactic CRs. The fantastic sensitivity of current WFoV instruments, and in particular LHAASO, has stirred the field, unveiling a real population of PeV sources, in a regime in which only a few sources were expected. This has also raised new exciting questions, opening the UHE range to new scientific quests.

\backmatter

\bmhead{Acknowledgments}

This article is the result of fruitful discussions during the 2nd HONEST (Hot Topics in High Energy Astrophysics) Workshop ``PeVatrons and their environments'' (see https://indico.desy.de/event/34265/). We would first and foremost thank the Scientific Organizing Committee of the Workshop, composed by the paper authors plus Tony Bell (University of Oxford), David Berge (DESY Zeuthen), Juan Cortina (CIEMAT), Petra Huentemeyer (Michigan Tech University), Sarah Recchia (INFN Torino) and Roberta Zanin (CTA Observatory); the Local Organizing Committee: Julia Eckert, Sonal Patel, Jonas Kramer, Mar\'ia Isabel Bernardos and to all the participants of the workshop without whom the fruitful discussions would not have been possible.

\section*{Declarations}
\begin{itemize}
\item Funding

R.L.-C. acknowledges the Ram\'on y Cajal program through grant RYC-2020-028639-I. He also acknowledges financial support from the Spanish "Ministerio de Ciencia e Innovaci\'on" (MCIN/AEI/10.13039/501100011033) through the Center of Excellence Severo Ochoa award for the Instituto de Astrofísica de Andalucía-CSIC (CEX2021-001131-S), and through grants PID2019-107847RB-C44 and PID2022-139117NB-C44.
E.A. acknowledges support by INAF under grant PRIN-INAF 2019, and by the European Union - Next Generation EU through grant PRIN-MUR 2022TJW4EJ. 
S.G. acknowledges support from Agence Nationale de la Recherche (project CRitiLISM, ANR-21-CE31-0028).
\item Competing interests

The authors declare no competing interests.
\item Ethics approval

Not applicable
\item Consent to participate

Not applicable
\item Consent for publication

Not applicable
\item Availability of data and materials

Not applicable
\item Code availability

Not applicable
\item Authors' contributions

EdOW and RL-C coordinated the manuscript writing. All authors meet the journal's authorship criteria and have reviewed, discussed, and commented on the review content. 

\end{itemize}






\bibliography{references}

\end{document}